\documentclass[prl,amssymb,twocolumn,showpacs,superscriptaddress,floatfix]{revtex4}

\usepackage{graphicx}
\usepackage{bm}
\usepackage{mathptmx}

\begin{document}

\title{Random sequential adsorption of straight rigid rods on a simple cubic lattice}

\author{G. D. Garc\'{\i}a}
\affiliation{Universidad Tecnol\'ogica Nacional, Facultad Regional San Rafael, \\
 Gral J. J. De Urquiza 340, C.P. M5602GCH, San Rafael, Mendoza, Argentina \\}

\author{F. O. Sanchez-Varretti}
\affiliation{Universidad Tecnol\'ogica Nacional, Facultad Regional San Rafael, \\
 Gral J. J. De Urquiza 340, C.P. M5602GCH, San Rafael, Mendoza, Argentina \\}

\author{P. M. Centres}
\affiliation{Departamento de F\'{\i}sica, Instituto de F\'{\i}sica Aplicada, Universidad Nacional de San Luis-CONICET, Ej\'ercito de los Andes 950, D5700HHW, San  Luis, Argentina}

\author{ A. J. Ramirez-Pastor$^{\dag}$}
\affiliation{Departamento de F\'{\i}sica, Instituto de F\'{\i}sica Aplicada, Universidad Nacional de San Luis-CONICET, Ej\'ercito de los Andes 950, D5700HHW, San  Luis, Argentina}

\begin{abstract}
Random sequential adsorption of straight rigid rods of length $k$ ($k$-mers) on a simple cubic lattice has been studied by numerical simulations and finite-size scaling analysis. The calculations were performed by using a new theoretical scheme, whose accuracy was verified by comparison with rigorous analytical data. The results, obtained for \textit{k} ranging from 2 to 64, revealed that (i) in the case of dimers ($k=2$), the jamming coverage is $\theta_j=0.918388(16)$. Our estimate of $\theta_j$ differs significantly from the previously reported value of $\theta_j=0.799(2)$ [Y. Y. Tarasevich and V. A. Cherkasova, Eur. Phys. J. B \textbf{60}, 97 (2007)]; (ii) $\theta_j$ exhibits a decreasing function when it is plotted in terms of the $k$-mer size, being $\theta_j (\infty)= 0.4045(19)$ the value of the limit coverage for large $k$'s; and (iii) the ratio between percolation threshold and jamming coverage shows a non-universal behavior, monotonically decreasing with increasing $k$.
\end{abstract}

\date{\today}
\maketitle

 %%%%%%%%%%%%%%%%%%%%%%%%%%%%%%%%%%%%%%%%%%%%%%%%%%%%%%%%%%%%%%%%%%%%%
 \section{Introduction}
 %%%%%%%%%%%%%%%%%%%%%%%%%%%%%%%%%%%%%%%%%%%%%%%%%%%%%%%%%%%%%%%%%%%%%

The study of systems of hard non-spherical particles is one of the central problems in statistical mechanics and has been attracting a great deal of interest since long ago. In a seminal contribution to this subject, Onsager \cite{Onsager} predicted that rod-like particles interacting with only excluded-volume interaction can lead to long-range orientational (nematic) order. This nematic phase, characterized by a big domain of parallel molecules, is separated from an isotropic state by a phase transition occurring at a finite critical density. Numerous model systems of anisotropic objects, ranging from inorganic to organic and also biological particles, were analyzed by using the Onsager's theory. An extensive overview of this field can be found in the excellent reviews of Refs. \cite{Stroobants,Gabriel}.

The long-range orientational order disappears in the case of irreversible adsorption (no desorption), where the
distribution of adsorbed objects is different from that obtained at equilibrium. Thus, at high coverage, the
equilibrium state corresponds to a nematic phase with long-range correlations, whereas the final state generated by irreversible adsorption is a disordered state (known as jamming state), in which no more objects can be deposited due to the absence of free space of appropriate size and shape. The jamming state has infinite memory of the process and the orientational order is purely local \cite{Evans,Privman,Talbot}.

The phenomenon of jamming plays an important role in numerous systems where the deposition of objects is irreversible over time scales of physical interest \cite{Erban}. The random sequential adsorption (RSA), introduced by Feder \cite{Feder}, has served as a paradigm for modeling irreversible deposition processes. The main features of the RSA model are: (1) the objects are put on randomly chosen sites, (2) the adsorption is irreversible and (3) at
any time only one object is being adsorbed, so that the process takes place sequentially. In this theoretical frame, anisotropic particles of different shapes and sizes have been studied: linear $k$-mers (particles occupying $k$ adjacent lattice sites) \cite{Redner,Bonnier,Vandewalle,Kondrat,Lebovka}, flexible polymers \cite{Paw1,Paw2}, T-shaped objects and crosses \cite{Adam}, squares \cite{Nakamura,Vigil,Dickman,Kriu}, disks \cite{Connelly}, regular and star polygons \cite{Ciesla}, etc. In all cases, the limiting or jamming coverage, $\theta_j$, strongly depends on the shape and size of the depositing particles.

In the case of lattice models straight rigid rods (or linear $k$-mers), which is the topic of this paper, the RSA problem has been exactly solved for one-dimensional (1D) substrates. In this limit, $\theta(t)$ can be written as \cite{Redner},
\begin{equation}
\theta(t) = k \int_0^t \exp \left[ -u-2 \sum_{j=1}^{k-1} \left( \frac{1-e^{-ju}}{j} \right) \right]du.    \label{t1D}
\end{equation}
From Eq. (\ref{t1D}), the dependence on $k$ of the jamming coverage can be obtained. Note that $\theta(t)$ represents the fraction of lattice sites covered at time $t$ by the deposited objects and, consequently, $\theta(t=\infty)=\theta_j$.

In two and three dimensions, the inherent complexity of the system still represents a major difficulty to the development of accurate analytical solutions, and  computer simulations appear as a very important tool for studying this subject. In this direction, Bonnier et al. investigated the deposition of linear $k$-mers (with $k$ ranging between 2 and 512 on a two-dimensional (2D) square lattice \cite{Bonnier}. The authors found that the jamming concentration monotonically decreases and tends to $0.660(2)$ as the length of the rods increases.

Vandewalle et al. studied the percolation and jamming phenomena for straight rigid rods of size $k$ on 2D square lattices \cite{Vandewalle} and found that, for values of $k$ between 2 and 10, the ratio $\theta_p/\theta_j$ (being $\theta_p$ the percolation threshold) remains constant $\theta_p/\theta_j \approx 0.62$, regardless of the length of the particle. Based on this finding, the authors suggested that both critical phenomena (percolation and jamming) are intimately related. Kondrat and P\c{e}kalski \cite{Kondrat} extended the study of Ref. \cite{Vandewalle} to larger lattices (lattice size $L=30, 100, 300, 1000$ and $2500$) and longer objects ($1 \leq k \leq 2000$). The results obtained revealed that: (1) as reported in Ref. \cite{Bonnier}, the jamming coverage decreases monotonically approaching the asymptotic value of $\theta_j=0.66(1)$ for large values of $k$; (2) the percolation threshold is a nonmonotonic function of the size $k$, showing a minimum around $k=13$; and (3) the ratio of the two thresholds
 $\theta_p/\theta_j$ is nonmonotonic too: after initial growth, it stabilizes between $k=3$ and $k=7$, and then it grows logarithmically.

Recently, Lebovka et al. studied an anisotropic RSA of straight rigid rods on 2D square lattices \cite{Lebovka}. In this model, the vertical and horizontal orientations occur with different probabilities, and the degree of anisotropy of the system can be characterized by an order parameter measuring the difference (normalized) between the number of line segments oriented in the vertical direction and the number of line segments oriented in the horizontal direction. 
The authors investigated the effect of $k$-mer alignment on the jamming properties. In the case of isotropic systems (order parameter equal to zero), the results obtained by Lebovka et al. are in excellent agreement with previous simulation data in Ref. \cite{Kondrat}.

The RSA problem becomes more difficult to solve when the objects are deposited on three-dimensional (3D) lattices, and only very moderate progress has been reported so far. In the line of present work, Tarasevich and Cherkasova examined   the percolation and jamming properties of dimers (straight rigid rods with $k=2$) on 3D simple cubic lattices and found that (1) $\theta_p=0.2555(1)$ and $\theta_j=0.799(2)$, and (2) in  contrast with percolation and jamming on square lattices, $\theta_p/\theta_j \approx 0.32$ for cubic lattices \cite{Tarasevich}. 

In the present paper, the study of Tarasevich and Cherkasova is extended to larger $k$-mers ($2 \leq k \leq 64$). Using computational simulations and a new method introduced here, the jamming properties of straight rigid rods on 3D simple cubic lattices are studied. The method is based on the calculation of the probability $W_L(\theta)$ that a lattice composed of $L \times L \times L$ elements reaches a coverage $\theta$. As it will be discussed in details in the next section, the intersection point of the curves $W_L(\theta)$ for different values of $L$ gives an accurate estimation of the jamming threshold in the infinite system. The main objectives of the paper are (1) to evaluate the accuracy and applicability of the new technique to estimate jamming properties, (2) to determine the dependence of the jamming coverage on the size of the deposited $k$-mers, and (3) to compare our results with the corresponding ones in Ref. \cite{Tarasevich}. The work is a natural extension of our previous research in the area of percolation and jamming of linear $k$-mers.   

The outline of the paper is as follows. In Sec. II we describe the deposition model, along with the simulation scheme and the procedure used to calculate the jamming coverage. In Sec. III we present the results and discussion of the simulations. Finally, the general conclusions are given in Sec. IV.

\section{Model, basic definitions and simulation scheme}

The following scheme is usually called standard model of deposition or RSA. Let us consider the substrate is represented by a three-dimensional simple cubic lattice of $M=L \times L \times L$ sites and periodic boundary conditions. In the filling process, straight rigid rods of length $k$ (with $k \geq 2$) are deposited randomly, sequentially, and irreversibly on an initially empty lattice. The procedure of
deposition is as follows: $(i)$ one of the three possible lattice directions $(x,y,z)$ and a starting site are randomly chosen; $(ii)$ if, beginning at the chosen site, there are $k$ consecutive empty sites, then a $k$-mer is deposited on those sites. Otherwise, the attempt is rejected. When $N$ rods are deposited, the concentration is $\theta = kN/M$.

Due to the blocking of the lattice by the already randomly deposited dimers, the limiting or jamming coverage, $\theta_j \equiv  \theta(t=\infty)$ is less than that corresponding to the close packing ($\theta_j < 1$). Note that $\theta(t)$ represents the fraction of lattice sites covered at time $t$ by the deposited objects. 
Consequently, $\theta$ ranges from 0 to $\theta_j$ for objects occupying more than one site, and the jamming coverage depends on the size of the deposited object \cite{Evans,Redner,Bonnier,Vandewalle,Kondrat,Lebovka}.

It is well known that it is a quite difficult matter to analytically determine the value of the jamming coverage for a given lattice. For some special types of lattices, geometrical considerations enable to derive their jamming thresholds exactly \cite{Redner}. For systems which do not present such a topological advantage, jamming
properties have to be estimated numerically by means of computer simulations.

\begin{figure}
\includegraphics[width=7.5cm,clip=true]{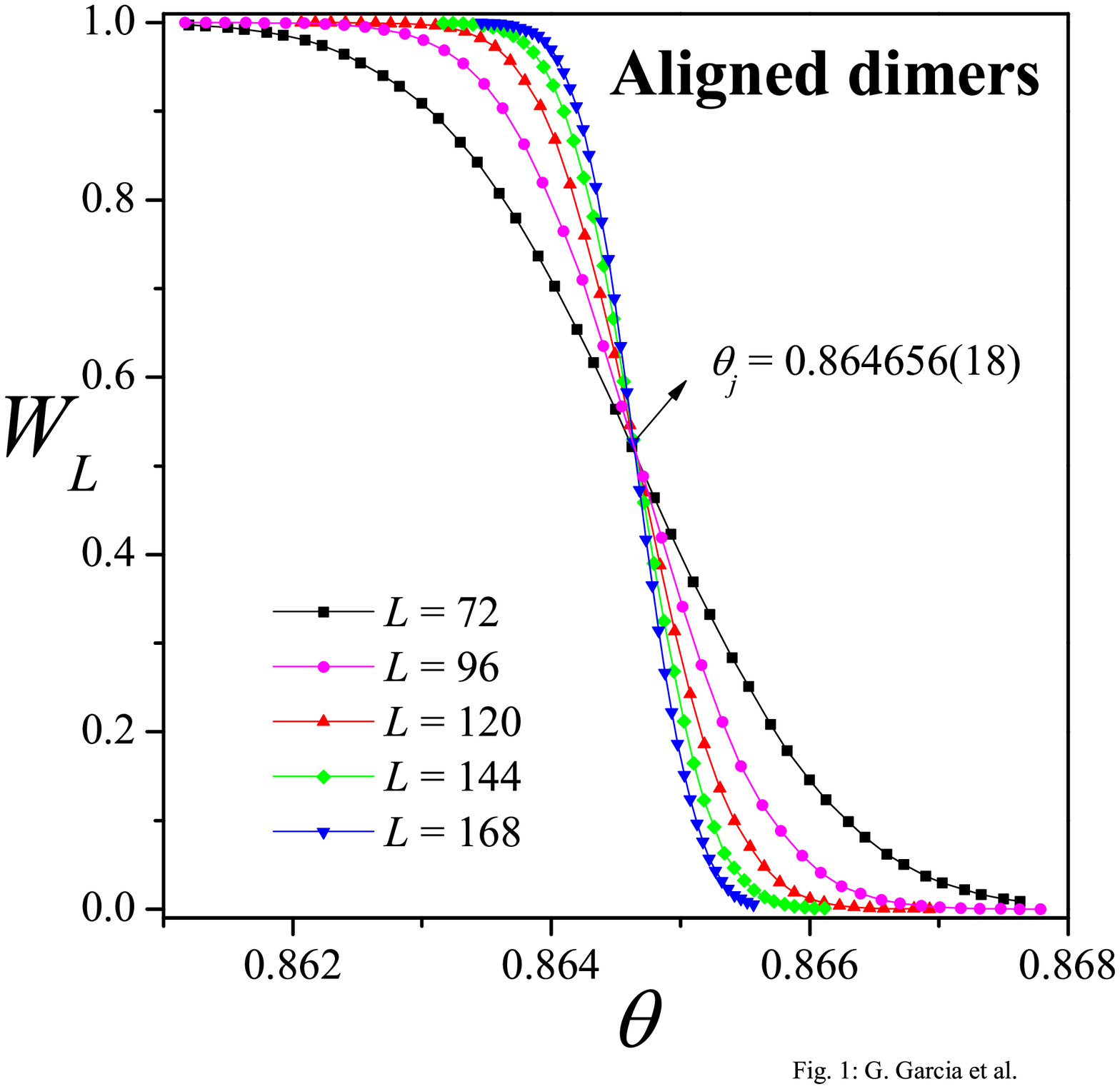}
\caption{\label{figure1} Curves of $W_L$ vs $\theta$ for fully aligned dimers on simple cubic lattices of different sizes as indicated. From their intersections one obtained $\theta_j$. In this case, $\theta_j=0.864656(18)$.}
\end{figure}

In order to calculate the jamming thresholds, we introduce the probability $W_L(\theta)$ that a lattice composed of $L \times L \times L$ elements reaches a coverage $\theta$. In the simulations, the procedure to determine $W_L(\theta)$ consists of the following steps: (a) the construction of the $M=L \times L \times L$ lattice (initially empty) and (b) the deposition of particles on the lattice up to the jamming limit $\theta_j$. In the late step, the quantity $m_i(\theta)$ is calculated as
\begin{equation}
m_i(\theta)=\left\{ 
\begin{array}{cc}
1 & {\rm for}\ \ \theta \leq \theta_j \\ 
0  & {\rm for}\ \ \theta > \theta_j .
\end{array}
\right. 
\end{equation}
$n$ runs of such two steps (a)-(b) are carried out for obtaining the number $m(\theta)$ of them for which a lattice  reaches a coverage $\theta$,
\begin{equation}\label{m}
m(\theta) = \sum_{i=1}^n m_i(\theta).
\end{equation}
Then, $W_L(\theta)=m(\theta)/n$ is defined and the procedure is repeated for different values of lattice sizes.
For infinite systems ($L \rightarrow \infty$), $W_L(\theta)$ is a step function, being 1 for $\theta \leq \theta_j$ and 0 for $\theta > \theta_j$. For finite values of $L$, $W_L(\theta)$ varies continuously between 1 and 0, with a sharp fall around $\theta_j$. In conclusion, $W_L(\theta)$ appears as a good parameter evidencing the jamming threshold and its definition is computationally convenient.

Using a system of fully aligned rods deposited on 3D cubic lattices as an example of application \footnote{In this case, the $k$-mers are deposited along one of the directions of the lattice, forming a nematic phase.}, we show in the following how to determine the corresponding jamming coverage from the probabilities $W_L(\theta)$. This model of fully aligned rods can be exactly solved \cite{Redner}, which allows us not only to illustrate the use of the method introduced here, but also to test its validity and accuracy by comparing our simulation results with those of Ref. \cite{Redner}. Five $k$-mer sizes were chosen for the comparison $k(=2,3,4,5,6)$. Then, a set of $n = 10 ^5$ runs were carried out for each pair $k$ and $L/k(=18,24,30,36,42)$. 

\begin{center}
Table I: Simulation and theoretical values of $\theta_j$ (as indicated in the text) for fully aligned rods on a simple cubic lattice and $k$ ranging from 2 to 6.
\end{center}
$$
\begin{array}{|c||c|c|}
\hline  \hline
 &  \multicolumn{2}{|c|}{\rm Aligned \ \rm rods}   \\
 \hline
k &  \theta_j \ \ [{\rm Eq.} (1)] & \theta_j \ \ [ W_L{\rm's}] \\
\hline
2 &   0.864665 & 0.864656(18)    \\
\hline
3 &   0.823653 & 0.823646(16)    \\
\hline
4 &  0.803893 & 0.803892(12)    \\
\hline
5  & 0.792276 & 0.792284(15)    \\
\hline
6 & 0.784630 & 0.784637(16)    \\
\hline
\end{array}
$$

In Fig. 1, the probabilities $W_L(\theta)$ are presented for a system of aligned dimers ($k=2$) and several values of $L$ as indicated. As it can be observed from the figure, curves for different lattice sizes cross each other in a unique point, which is located at a very well defined value in the $\theta$-axes determining the jamming coverage for each $k$. The procedure in Fig. 1 was repeated for $k(=3,4,5,6)$ (data do not shown here for reasons of space) and the results are collected in the second column of Table I. 

\begin{figure}
\includegraphics[width=7.5cm,clip=true]{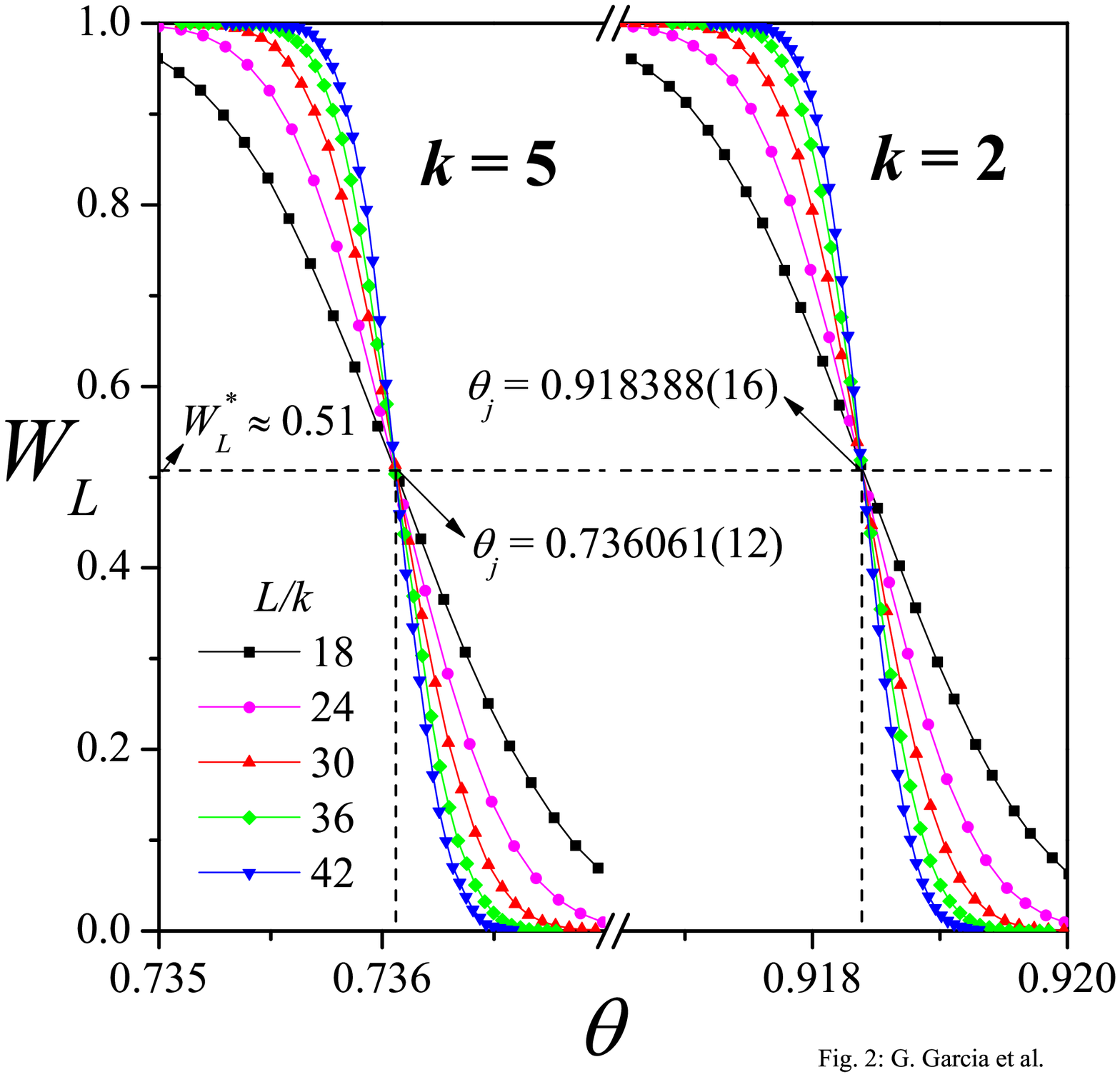}
\caption{\label{figure2} Curves of $W_L$ as a function of the density $\theta$ for two values of $k$-mer size ($k =$ 2, curves on the right, and $k =$ 5, curves on the left) and lattice sizes ranging between $L/k = 18$ and $L/k = 42$. Horizontal dashed line shows the $W^*_L$ point. Vertical dashed lines denote the jamming thresholds in the thermodynamic limit.}
\end{figure}

The high accuracy of the simulation data can be asserted by comparison with exact results for $\theta_j$ presented in Ref. \cite{Redner}. In fact, for a fully aligned system, as studied here, the $k$-mers are deposited only along an axis of the lattice, and the jamming problem reduces to the 1D case. In this limit, $\theta_j$ can be calculated from Eq. (\ref{t1D}) \cite{Redner}. The results are shown in the first column of Table I. A remarkable agreement is obtained between theoretical and simulation data, validating the applicability the method introduced here to calculate jamming thresholds.

In the next section, we will analyze the dependence of the jamming coverage on the size of the deposited objects for a system of straight rigid rods on a simple cubic lattice.

\section{Simulation Results: Random sequential adsorption of straight rigid rods on a simple cubic lattice}

The procedure described in Section II was applied to calculate the jamming coverage corresponding to a system of straight rigid rods on a simple cubic lattice. $k$-mers of length $k=2 \dots 64$ were considered, and the finite-size study was performed for lattices ranging between $( 36 \times 36 \times 36 )$ and $(1152 \times 1152 \times 1152)$ sites. In addition, $n = 10 ^5$ runs were carried out for each pair $k$ and $L$. These values of the parameters allowed us to obtain very accurate measurements of the jamming thresholds, with errors less than $0.008 \% $ in all cases.

\begin{figure}
\includegraphics[width=6cm,clip=true]{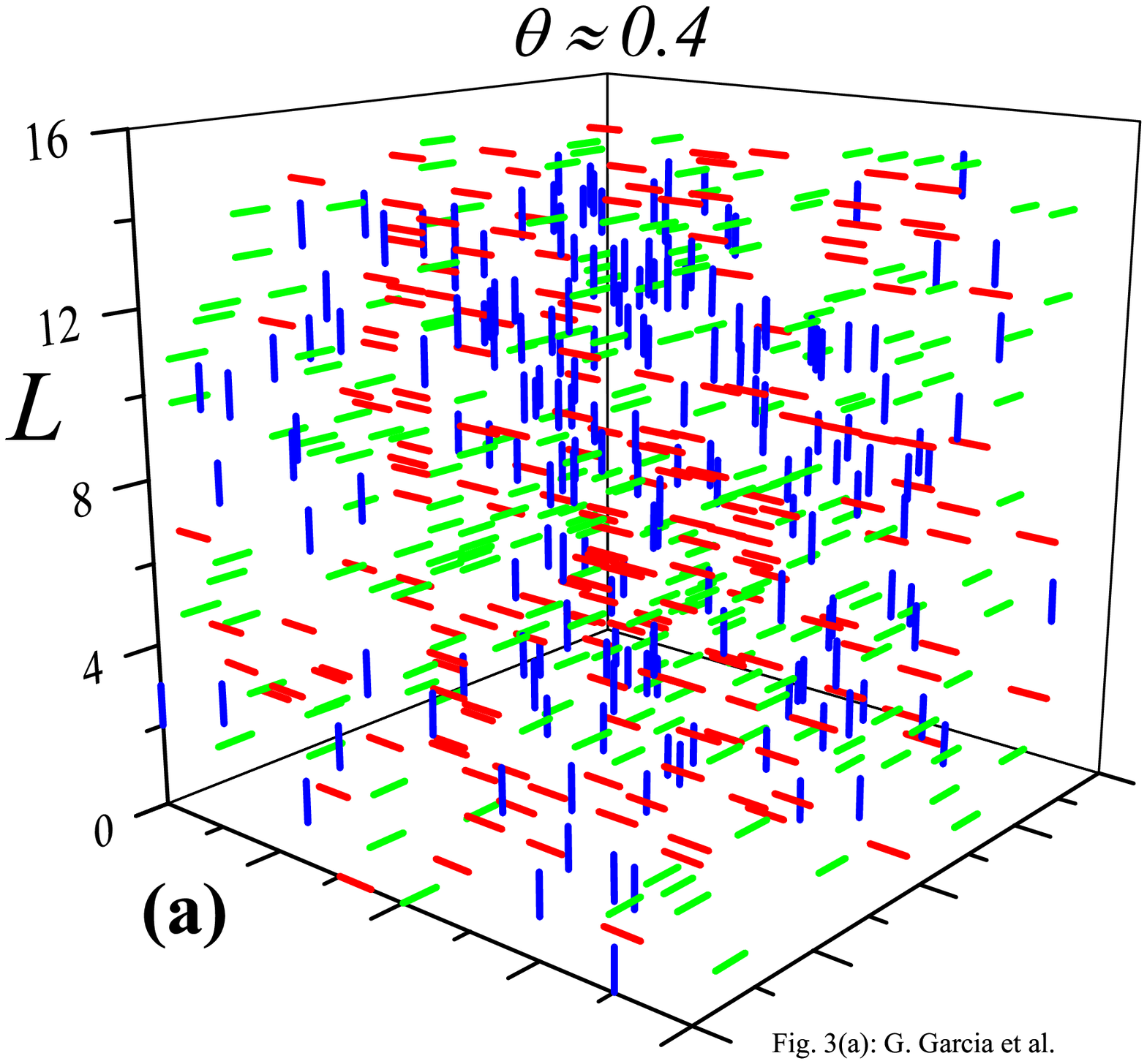}
\includegraphics[width=6cm,clip=true]{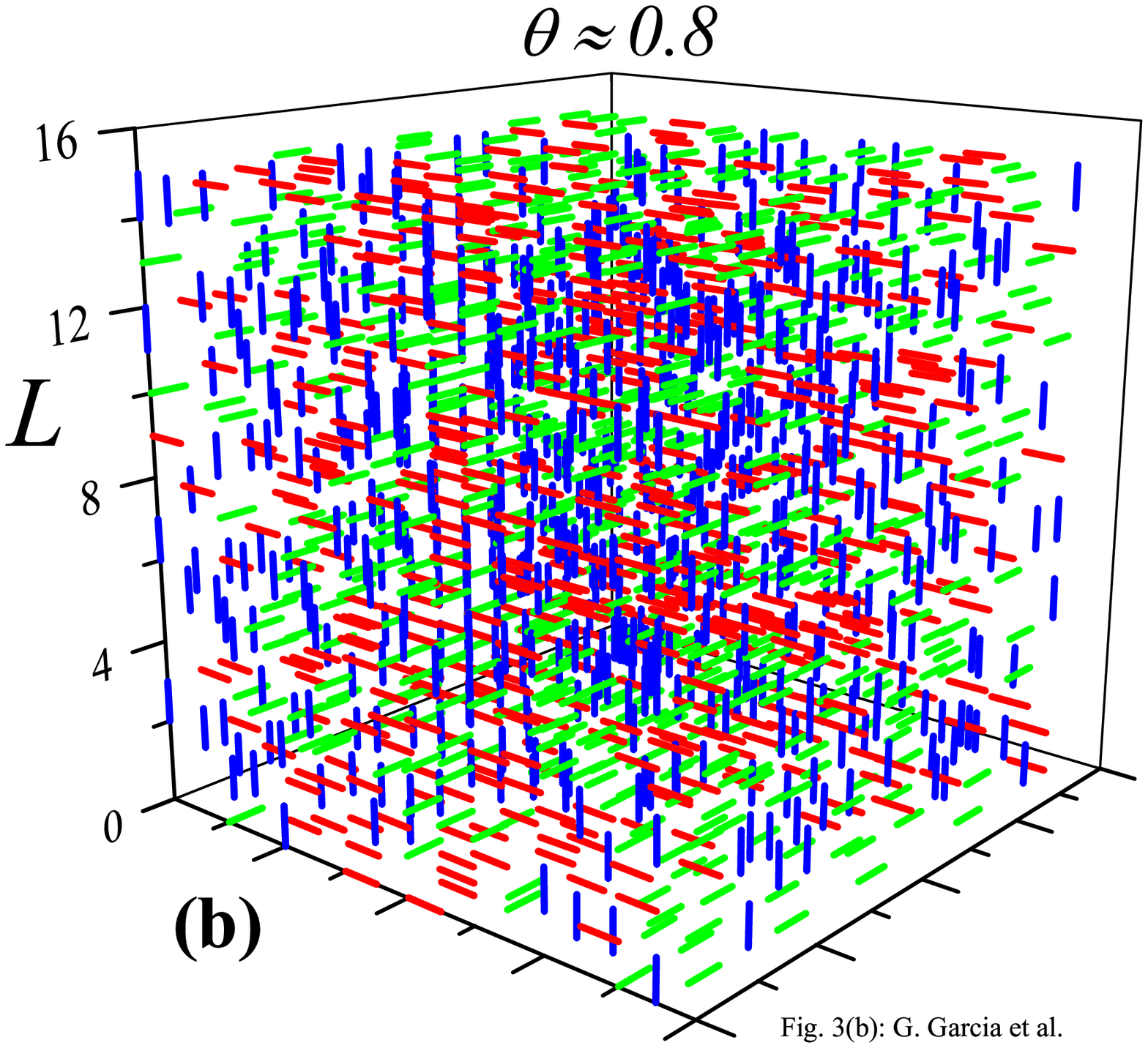}
\includegraphics[width=6cm,clip=true]{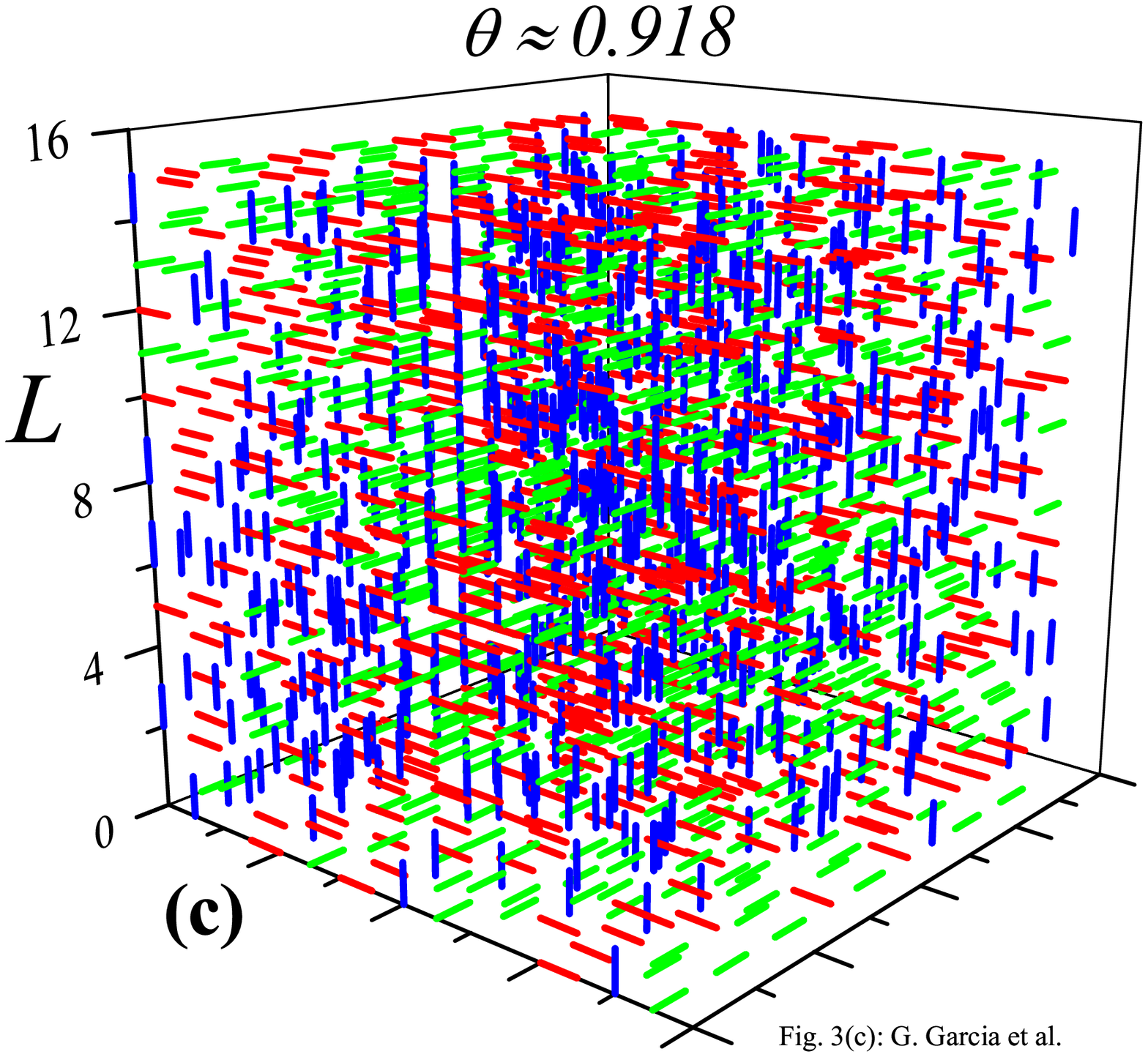}
\caption{\label{figure3} Typical configurations of dimers adsorbed on a $( 16 \times 16 \times 16 )$ lattice. (a) Low-density state ($\theta \approx 0.4$). (b) Intermediate-density state ($\theta \approx 0.8$).(c) Jamming state ($\theta \approx 0.918$). Different colors represent different directions.}
\end{figure}

In Fig. 2, the probabilities $W_L(\theta)$ are shown for two different values of $k$ ($k=2$ and $k=5$ as indicated) and $L/k(=18,24,30,36,42)$. From a simple inspection of the figure (and from data do not shown here for a sake of clarity) it is observed that: (a) for each $k$, the curves cross each other in a unique point $W^*_L$; (b) those points do not modify their numerical value for the different $k$ used (ranged between $k=2$ to $k=64$). In the cases of Fig. 2, $W^*_L=0.509(9)$ for $k=2$, and $W^*_L=0.505(9)$ for $k=5$; (c) those points are located at very well defined values in the $\theta$-axes determining the jamming threshold for each $k$, being $\theta_{j}=0.918388(16)$ for $k=2$, and $\theta_{j}=0.736061(12)$ for $k=5$; and (d) $\theta_{j}$ decreases for increasing $k$-mer sizes.

\begin{figure}
\includegraphics[width=7.5cm,clip=true]{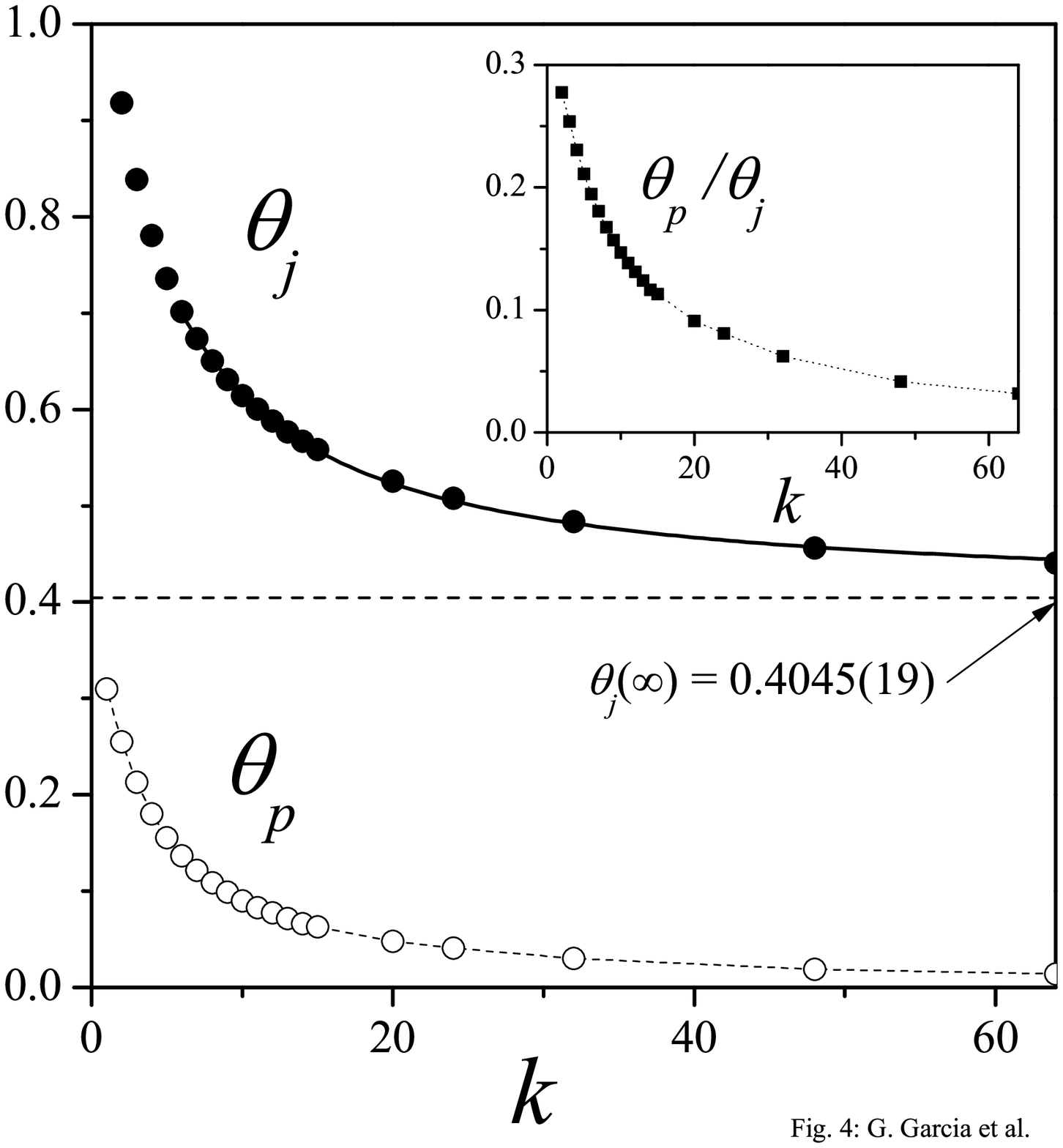}
\caption{\label{figure4} The thresholds $\theta_j$ (solid circles) and $\theta_p$ (open circles) as a function of $k$. The solid line corresponds of the best fit of the jamming values as indicated in the text. Inset: Ratio $\theta_p/\theta_j$ as a function of $k$.}
\end{figure}

In the case of dimers ($k=2$), the value of $\theta_{j}$ obtained in Fig. 2 differs significantly from the value $\theta_j=0.799(2)$ reported by Tarasevich and Cherkasova \cite{Tarasevich}. Due to the methodology used in this contribution, our estimate of $\theta_j$ is expected to be more accurate than that reported previously. In order to reinforce the validity of our result, typical configurations of dimers adsorbed on a simple cubic lattice are shown in Fig. 3: (a) low-density state ($\theta \approx 0.4$), (b) intermediate-density state ($\theta \approx 0.8$), and (c) jamming state ($\theta \approx 0.918$). Different colors represent different directions. Clearly, the value of $\theta_j$ informed in Ref. \cite{Tarasevich} corresponds to an intermediate state and is far from the limit density of the system.

The procedure of Fig. 2 was repeated for $k$ ranging between 2 and 64, and the results are shown in Fig. 4 (solid circles) and second column of Table II. The points corresponding to $k =$32, 48 and 64 were calculated for three relatively small values of $L/k$ (12, 18, 24). The jamming coverage decreases upon increasing $k$. At the beginning, for small values of $k$, the curve rapidly decreases. However, it flatten out for larger values of $k$ and finally asymptotically converges toward a definite value as $k \rightarrow \infty$. 

Following a similar scheme to that proposed in Ref. \cite{Bonnier} for straight rigid rods on 2D square lattices, the curve of $\theta_j$ vs. $k$ was fitted to a function
\begin{equation}\label{tjinf}
\theta_j (k)= 0.4045(19) + \frac{2.63(5)}{k} - \frac{5.2(2)}{k^2}  \  \   \   \  \  (k \geq 6),
\end{equation}
being $\theta_j (\infty)= 0.4045(19)$ the result for the limit coverage of a simple cubic lattice by infinitely long $k$-mers.

Figure 4 includes also the behavior of the percolation threshold $\theta_p$ as a function of $k$ for the system studied here (open circles). The corresponding numerical values, obtained from Ref. \cite{Garcia}, are tabulated in the third column of Table II. The percolation threshold is a decreasing function of $k$. Combining the results of $\theta_j$ and $\theta_p$, the ratio $\theta_p/\theta_j$ was calculated (see inset in Fig. 4 and fourth column of Table II). As for 2D square lattices, the ratio between percolation and jamming thresholds exhibits a non-universal behavior. In this case, $\theta_p/\theta_j$ decreases in all range of $k$.

\section{Conclusions}

Random sequential adsorption of straight rigid $k$-mers deposited on a simple cubic lattice has been studied by numerical simulations and finite-size analysis. Several conclusions can be drawn from the present work. 

On the one hand, a new theoretical scheme to determine jamming thresholds was introduced here. The method relies on the definition of the probability $W_L(\theta)$ that a lattice composed of $L \times L \times L$ elements reaches a coverage $\theta$. The value of $\theta_j$ can be obtained from the crossing point of the curves of $W_L$ for different lattice sizes. Comparisons with rigorous analytical results demonstrated the method's accuracy.

\begin{center}
Table II: Values of $\theta_j$, $\theta_p$ and $\theta_p/\theta_j$ (as indicated in the text) for isotropic rods on a simple cubic lattice and $k$ ranging from 2 to 64.
\end{center}
$$
\begin{array}{|c||c|c|c|}
\hline  \hline
 & \multicolumn{3}{|c||}{\rm Isotropic \ \rm rods}   \\
 \hline
k & \theta_j \ \ [ W_L{\rm's}] & \theta_p \ \ {\rm Ref. \ \ [24]} & \theta_p/\theta_j  \\
\hline
2 & 0.918388(16) & 0.2555(1) \ \ {\rm  [23]}   & 0.2777    \\
\hline
3 & 0.838860(14) & 0.2129(1)  & 0.2538     \\
\hline
4 & 0.780344(16) & 0.1800(1) & 0.2307     \\
\hline
5 & 0.736061(12) & 0.1555(1) & 0.2113     \\
\hline
6 & 0.701346(13) & 0.1364(1)  & 0.1945     \\
\hline
7 & 0.673355(14) & 0.1218(1) & 0.1809     \\
\hline
8 & 0.650282(11) & 0.1089(1)  & 0.1675     \\
\hline
9 & 0.630901(12) & 0.0990(1) & 0.1569     \\
\hline
10 & 0.614384(15) & 0.0901(1) & 0.1467     \\
\hline
11 & 0.600130(10) & 0.0831(1)  & 0.1385     \\
\hline
12 & 0.587696(11) & 0.0772(1) & 0.1314     \\
\hline
13 & 0.576780(14) & 0.0714(1)  & 0.1238     \\
\hline
14 & 0.567044(11) & 0.0661(1) & 0.1164     \\
\hline
15 & 0.558360(13) & 0.0632(1) & 0.1128     \\
\hline
20 & 0.525676(18) & 0.0478(1)  & 0.0909     \\
\hline
24 & 0.507750(18) & 0.0411(1) & 0.0807     \\
\hline
32 & 0.483360(34) & 0.0299(1)  & 0.0620     \\
\hline
48 & 0.456071(32) & 0.0191(1) & 0.0417     \\
\hline
64 & 0.440655(34) & 0.0143(1) & 0.0318     \\
\hline
\end{array}
$$

On the other hand, the jamming coverage dependence on the $k$-mer size has been reported. In the case of dimers ($k=2$), $\theta_j=0.918388(16)$ was obtained. This value differs significantly from the value $\theta_j=0.799(2)$ calculated by Tarasevich and Cherkasova \cite{Tarasevich}. Due to the methodology used in this contribution, our estimate of $\theta_j$ is expected to be more accurate than that reported previously.

On the basis of the behavior of $\theta_j$ vs. $k$ in the range $2 \leq k \leq 64$, and the best fit to this curve, the expression $\theta_j (k)= 0.4045 + 2.63/k - 5.2/k^2$ was found, being $\theta_j (\infty)= 0.4045(19)$ the 
result for the limit coverage of a simple cubic lattice by infinitely long $k$-mers.

Finally, and based on previous findings from our group \cite{Garcia}, the possible relationship between percolation threshold and jamming coverage was examined. The results showed a non-universal behavior for the ratio $\theta_p/\theta_j$, which decreases upon increasing $k$.

\section{ACKNOWLEDGMENTS}

This work was supported in part by CONICET (Argentina) under project number PIP 112-201101-00615; Universidad Nacional de San Luis (Argentina) under project 322000; Universidad Tecnol\'ogica Nacional, Facultad Regional San Rafael, under project PID UTN 1835 Disp. 284/12; and the National Agency of Scientific and Technological Promotion (Argentina) under project PICT-2013-1678.

\end{document}